\RequirePackage{fix-cm}
\documentclass[smallextended]{svjour3}       
\smartqed  
\usepackage{graphicx}
\usepackage{amsmath}
 \usepackage{mathptmx}      
\usepackage{latexsym}
\usepackage{physics}
\newcommand{\Beq}{\begin{equation}}
\newcommand{\Eeq}{\end{equation}}
\renewcommand{\vec}[1]{\mathrm{#1}}

\begin{document}

\title{Calculation of Relativistic Single-Particle States
}


\author{  D.~Wingard
\and   B.~K\'onya
\and Z.~Papp$^{*}$
}


\institute{  D.~Wingard
	\at
              Department of Physics  and Astronomy \\
              California State University Long Beach \\
              Long Beach, California, USA        
\and
            B.~K\'onya
	\at
              Department of Physics \\
              Lund University, Lund, Sweden
 \and             
                Z.~Papp$^{*}$
	\at
              Department of Physics  and Astronomy \\
              California State University Long Beach \\
              Long Beach, California, USA\\
              \email{Zoltan.Papp@csulb.edu} 	      
}

\date{Received: date / Accepted: date}

\maketitle

\begin{abstract}
 
A computational method is proposed to calculate bound and resonant states by solving the 
Klein-Gordon and Dirac equations for real and complex energies, respectively.
The method is an extension of a non-relativistic one, where the potential is represented in a Coulomb-Sturmian basis. This basis  facilitates the
exact analytic evaluation of the Coulomb Green's operator in terms of a continued fraction. In the extension to relativistic problems,
we cast the Klein-Gordon and Dirac equations into an effective Schr\"odinger form. Then the solution method is basically an
analytic continuation of non-relativistic quantities like the angular momentum, charge, energy and potential into the effective relativistic counterparts.

\end{abstract}

\section{Introduction }

The most often used theoretical tool for atomic and nuclear  physics is quantum mechanics, mostly its non-relativistic version. 
Usually, the effects of relativity are taken into account as the non-relativistic limit of the relativistic equations. 
While there are a large number of methods for non-relativistic calculations, methods for relativistic calculations are rather scarce.  

The aim of this work is to generalize an approximation method, that has been rather successful in non-relativistic quantum mechanics,
to relativistic calculations. We present a method that can equally solve the Schr\"odinger, the Klein-Gordon and the Dirac equations for 
bound and resonant states and for Coulomb plus short range potentials. 

The computational method has been developed a while ago \cite{papp1987bound,papp1988potential,PhysRevC.61.034302} and 
has been applied for solving various problems in nuclear and atomic physics 
like the Faddeev equation with Coulomb-like interactions \cite{PhysRevC.54.50,PhysRevC.55.1080,PhysRevA.63.062721}.

In this work, we want to develop a computational method that will allow us to 
incorporate relativistic quantum mechanics in our studies.
The design of this work is as follows. In Section \ref{sec2} we review  the method applied to the Schr\"odinger equation.
Then, in Sections \ref{sec3} and \ref{sec4} we show how to extend it for solving the Klein-Gordon and the Dirac equations, respectively. 
In Section \ref{sec5} we present some numerical illustrations and in Section \ref{sec6} we summarize our findings.

\section{Solution of the Schr\"odinger equation}
\label{sec2}

We consider a Hamiltonian with a Coulomb   plus short-range  potential in angular momentum
channel $l$ 
\begin{equation}
h_{l} = h_{l}^{0} + Z /r + v_{l}^{(s)},
\end{equation}
where $h_{l}^{0}$ is the non-relativistic kinetic energy operator, $Z$ is the charge number, 
and $v^{(s)}$ is a short-range potential.
This Hamiltonian gives us the 
 Schr\"odinger eigenvalue equation
\begin{equation}
( h_{l}^{0}  + Z/r + v_{l}^{(s)} )  | \psi_{l} \rangle = E  | \psi_{l} \rangle.
\end{equation}
If  we represent the momentum operator by a derivative in terms of the spatial variable, the eigenvalue problem becomes
 a differential equation, which can be solved with the appropriate boundary conditions.  

We can also cast the Schr\"odinger eigenvalue equation into a Lippmann-Schwinger form. 
If we are concerned about bound 
and resonant states, we need to solve the homogenous Lippmann-Schwinger equation
\begin{equation}\label{lssc}
|\psi_{l} \rangle =   g_{l}^{C}(Z,E  ) v_{l}^{(s)}|\psi_{l}  \rangle~,
\end{equation}
with negative real and positive complex  energies, respectively. 
Here 
\begin{equation}
g_{l}^{C}(Z,E)=(E-h_{l}^{0}-Z/r)^{-1}
\end{equation}
is the  Coulomb Green's operator.

We solve the Lippmann-Schwinger equation by approximating the short-range potential $v_{l}^{(s)}$ on a Hilbert-space
basis. For that purpose we take the 
Coulomb-Sturmian (CS) basis.  
The CS functions in angular momentum $l$ are defined by
\begin{equation}
\langle r | n l \rangle = \left( \frac{{\Gamma(n+1)}}{{\Gamma(n+2l+2)}}\right)^{1/2} \exp(-br) (2br)^{l+1} L_{n}^{2l+1}(2br)~,
\label{csfun}
\end{equation}
where $n=0,1,2, \dots$, $L$ is the Laguerre  polynomial and $b$ is a parameter. Together with 
$
\langle r | \widetilde{nl} \rangle = \langle r |  {nl} \rangle /r
$
these functions are orthonormal
$
\langle nl | \widetilde{n' l} \rangle = \delta_{n n'}
$
and form a complete set
$\lim_{N\to\infty}\sum_{n=0}^{N} | nl \rangle \langle \widetilde{n' l} | = 1$.

The finite dimensional representation of the short-range potential is given by
\begin{equation}\label{vapprox}
v_{l}^{(s)} \approx \sum_{n n'}^{N} | \widetilde{nl}\rangle  \underline{\tilde{v}}^{(s)}_{l,n n'}  \langle \widetilde{n' l} |~.
\end{equation}
To construct the matrix $\underline{\tilde{v}}^{(s)}_{l,n n'}$ 
we calculate the matrix elements  $\underline{v}^{(N')} = \langle{nl} | v_{l}^{(s)} | n' l \rangle$ up to $N' \ge N$, numerically in general, 
invert $\underline{v}^{(N')}$, then truncate the $N' \times N'$ inverse matrix to a $N\times N$ matrix and invert again 
to obtain the $N\times N$ matrix $\underline{\tilde{v}}^{(s)}_{l,n n'}$  \cite{brown2013approximations}.

With this approximation, the Lippmann-Schwinger equation (\ref{lssc}) becomes 
\Beq
|\psi_{l} \rangle =  \sum_{n n'}^{N} g_{l}^{C}(Z,E  )    | \widetilde{nl}\rangle  \underline{\tilde{v}}^{(s)}_{l,n n'}  \langle \widetilde{n' l} |   \psi_{l}  \rangle.
\label{eqls44}
\Eeq
We can see that  $|\psi_{l} \rangle$ is determined by the coefficients $\langle \widetilde{n' l} |   \psi_{l}  \rangle$ where $n'$ goes only up to $N$. 
Therefore, to determine these coefficients we multiply from the left by $\bra{ \widetilde{n'' l} }$ with $n''$ up to $N$ as well. This
results in a matrix
equation for the CS coefficients of the wave function $\underline{\psi} =  \langle \widetilde{nl} | \psi_{l} \rangle$
\begin{equation}\label{lsscm}
\underline{\psi}_{l}  =   \underline{g}_{l}^{C}(Z,E)  \underline{\tilde{v}}_{l}^{(s)} \underline{\psi}_{l},
\end{equation}
 where 
\begin{equation}
  \underline{g}_{l}^{C}(Z,E)  = \langle \widetilde{nl} | g_{l}^{C}(Z,E) | \widetilde{n'l}\rangle.
\end{equation}
The equation is solvable if the determinant is zero
\begin{equation}
|( \underline{g}_{l}^{C}(Z,E))^{-1}   - \underline{\tilde{v}}_{l}^{(s)} | =0.
\end{equation}

The calculation of the matrix $(\underline{g}_{l}^{C})^{-1}$ is based on the  infinite 
symmetric tridiagonal representation
 \Beq
 \mel{nl;b}{(z-\hat{h}_{l}^{(C)})}{n'l;b} = J_{n n'} =
 \begin{cases} 
 \displaystyle{\frac{k^{2}-b^{2}}{2m/\hbar^{2}\, b}} (n+l+1) -Z & \text{for} \quad  n' =n \\
  \displaystyle{ -\frac{k^{2} + b^{2}}{4m/\hbar^{2}\, b}  } \sqrt{(n+1)(n+2l+2) } & \text{for} \quad n' = n+1 \\
    \displaystyle{ -\frac{k^{2} + b^{2}}{4m/\hbar^{2}\, b}  } \sqrt{n(n+2l+1) } & \text{for} \quad n' = n-1 \\
    0 & \text{otherwise},  
 \end{cases}
 \Eeq
 where $k=\sqrt{2m/\hbar^{2}\, E}$.
It has been shown in Refs.\  \cite{Konya:1997JMP,PRADemir2006} that the $N\times N$ matrix $(\underline{g}_{l}^{C})^{-1}$
is   identical  to the $N\times N$ matrix $\underline{J}$ plus a correction term
in the bottom-right matrix element
\Beq
[\underline{g}_{l}^{C}(Z,E)]^{-1} =  \underline{J}_{l}^{C} - \delta_{i,N} \delta_{j,N} J_{N,N+1} C_{N+1} J_{N+1,N}.
\Eeq
 This correction is given in terms of $_{2}F_{1}$ hypergeometric functions
\Beq
 C_{N} = - \frac{ 4m /\hbar^{2} \, b}{(b-ik)^{2} (N+l+i \gamma) } 
 \frac{ \mbox{}_{2}F_{1} (-l+i\gamma,N+1;N+l+2+i\gamma; (b+ik )^{2}/(b-ik)^{2} )}{\mbox{}_{2}F_{1} (-l+i\gamma,N;N+l+1+i\gamma; (b+ik )^{2}/(b-ik)^{2} ) },
 \Eeq
 where $\gamma = Z m/k$. The ratio of hypergeometric functions with this combination of indexes can be evaluated by a continued fraction 
(see eq. Ref.\ \cite{baker1975essentials}).

In this approach the only approximation is the finite-basis representation of the short-range potential.  
As $N\to \infty$ the convergence is guaranteed, although the method is not variational. Only the short-range potential
is approximated, not the whole Hamiltonian. 
As a result, the convergence to the energy is not from above like in an usual Hilbert-space basis approximation scheme. 
If the parameter $b$ matches the size of the potential, the convergence is very fast.

The evaluation of the energy dependent 
$\underline{g}_{l}^{C}(Z,E)$  is exact and analytic, so the method can readily be extended to complex energies to calculate resonances, 
as has been shown in Refs.\ \cite{papp1987bound,Konya:1997JMP}.

In this method the solution is not a linear combination of basis functions. Rather, as Eq.\ (\ref{eqls44}) shows
\Beq
\braket{r}{\psi_{l}} = \sum_{n}^{N} c_{n} \,\mel*{r}{g_{l}^{C}(Z,E)}{\widetilde{nl}}, 
\Eeq
where $c_{n}= \sum_{n'}  \underline{\tilde{v}}^{(s)}_{l,n n'}  \braket*{ \widetilde{n' l} }{  \psi_{l} }$.
The Green's function in configuration space is given by the regular and irregular Coulomb functions $\varphi_{l}^{C}$ and $f_{l}^{C}$, respectively.
Therefore
\Beq
\mel*{r}{g_{l}^{C}(Z,E)}{\widetilde{nl}} \sim \int_{0}^{\infty} \dd{r'} \varphi_{l}^{C}(r_{<}) f_{l}^{C}(r_{>}) \braket*{r'}{\widetilde{nl}},
\Eeq
where $r_{<} = \min(r,r')$ and $r_{<} = \max(r,r')$. This integral behaves like $\phi_{l}^{C}(r)$ as $r\to 0$ and like $f_{l}^{C}(r)$ as $r\to \infty$,
which is  the correct Coulomb-like asymptotic behavior \cite{PhysRevA.46.4437},
 irrespectively of the basis parameter $b$ and the energy $E$. It should be noted however, that usually we don't need the 
 wave function, we need
 matrix elements of operators representing physical quantities. If an observable is represented by operator $O$, 
 we can approximate it on Hilbert space basis
 like in Eq.\ (\ref{vapprox})
 \Beq
 O \approx \sum_{n n'}^{N} | \widetilde{nl}\rangle  \underline{\tilde{O}}^{(s)}_{n n'}  \langle \widetilde{n' l} |.
 \Eeq
Then, the expectation value between eigenstates   reads
\Beq
\expval{O}{\psi} \approx \sum^{N}_{n m m' n'} c_{n} c_{n'} \underline{g}^{C}_{n m} \underline{\tilde{O}}_{m m'} \underline{g}^{C}_{m' n'}.
\Eeq

\section{Extension to the Klein-Gordon equation}
\label{sec3}

The relativistic spin-$0$ Klein-Gordon equation with potential term associated with the energy is given by
\begin{equation} 
\left( E -V \right)^{2} \psi  = {{p}}^{2} c^{2} \psi  +  {m^{2} c^{4}}    \psi.
\label{eq:56vp}
\end{equation}
If we divide  by $2mc^{2}$ we obtain
\Beq
\left[  \frac{1}{2m}  {{p}}^{2}  +\frac{mc^{2}}{2} - \frac{1}{2mc^{2}} ( E^{2} - 2EV +V^{2}) \right] \psi = 0.
\label{redeq}
\Eeq
By introducing the effective energy
\Beq
\varepsilon = \frac{1}{2mc^{2}} ( E^{2} -m^{2}c^{4}) 
\Eeq
and effective potential
\Beq
 \tilde{V} = V \left( \frac{E}{mc^{2}}   - \frac{V}{2mc^{2}} \right),
\Eeq
we can write the Klein-Gordon equation in a more familiar form
\Beq
 \tilde{H} \psi = \varepsilon \psi ,
\Eeq
where
\Beq
\tilde{H} = \frac{1}{2m}  {{p}}^{2}  + \tilde{V} .
\Eeq
If we separate off the rest energy, $E= mc^{2} + E'$, we find
\Beq
\varepsilon = E' \left( 1 + \frac{E'}{2mc^{2}} \right) 
\Eeq
and
\Beq
\tilde{V} = V \left( 1 +  \frac{E'}{mc^{2}}   - \frac{V}{2mc^{2}} \right).
\Eeq
We can see that in the non-relativistic limit $\epsilon \sim E'$ and $\tilde{V} \sim V$.

For spherical potentials,  the effective Hamiltonian commutes with the  angular momentum operators. 
Thus, $\{\tilde{H}, {L}^{2}, {L}_{z}\}$ form a 
complete set of commuting observables. Assuming a Coulomb plus short range potential again
\Beq
V(r)   = Z/r + v_{l}^{(s)}(r),
\Eeq
the effective Hamiltonian becomes
\Beq
\tilde{H}=  -\frac{\hbar^{2}}{2m}  \pdv[2]{r} + \frac{\hbar^{2}l(l+1)}{2mr^{2}}  + 
   \left(\frac{Z}{r} + v_{l}^{(s)} \right) \left(1 + \frac{E'}{mc^{2}}  -   \frac{1}{2mc^{2}}  \left(\frac{Z}{r} + v_{l}^{(s)} \right)  \right).
\Eeq
Reorganizing, we find
\Beq
\tilde{H} = -\frac{\hbar^{2}}{2m}  \pdv[2]{r} + \frac{\hbar^{2} }{2m}  \frac{ l(l+1) - Z^{2} \alpha^{2} }{r^{2}}   + \frac{Z'}{r} 
     + w_{l}^{(s)},
     \label{kgeffw}
\Eeq
where $\alpha=e^{2}/(\hbar c)$ is the fine structure constant with $e$ electric charge,  
\Beq
Z' = Z  \left(1 + \frac{E'}{mc^{2}} \right) 
\Eeq
and
\Beq
w_{l}^{(s)} = v_{l}^{(s)}  \left(1 + \frac{E'}{mc^{2}}  -   \frac{ 1}{mc^{2}}   \frac{ Z}{r}   -   \frac{ 1}{2mc^{2}}  v_{l}^{(s)}   \right).
\Eeq
If we equate  
\Beq
l(l+1) - Z^{2}\alpha^{2} =  \lambda ( \lambda+1),
\Eeq 
we find  that
\Beq
\lambda = -1/2 + \sqrt{(l+1/2)^{2}-Z^{2}\alpha^{2}}.
\Eeq
Then Eq.\ (\ref{kgeffw}) becomes
\Beq
 \left[ -\frac{\hbar^{2}}{2m}  \pdv[2]{r} + \frac{\hbar^{2} \lambda (\lambda+1)  }{2mr^{2}}   + \frac{Z'}{r} 
     + w_{l}^{(s)} \right]  \psi_{l}(r)  = \varepsilon  \psi_{l}(r).
\Eeq

This equation looks just like an ordinary radial Schr\"odinger equation with correspondences 
$l \to \lambda$, $Z\to Z'$ and $v_{l}^{(s)} \to w_{l}^{(s)}$. Then, the corresponding Lippmann-Schwinger equation reads
\begin{equation}\label{lsscmmm}
|\psi_{l} \rangle =   g_{\lambda}^{C}(Z',\varepsilon) w_{l}^{(s)}|\psi_{l}  \rangle~.
\end{equation}
To determine the solution we perform an analytic continuation in the CS basis 
$| nl \rangle \to | n \lambda \rangle$ and the determinant equation becomes
\begin{equation}
|( \underline{g}_{\lambda}^{C}(Z',\varepsilon))^{-1}   - \underline{\tilde{w}}_{l}^{(s)} | =0,
\end{equation}
where
\begin{equation}
\underline{g}_{\lambda}^{C}(Z',\varepsilon) =  \langle \widetilde{n \lambda } | g_{\lambda}^{C}(Z',\varepsilon) | \widetilde{n' \lambda}\rangle
\end{equation}
and
\begin{equation}
\underline{\tilde{w}}^{(s)} = \langle{n \lambda} | w_{l}^{(s)} | n' \lambda \rangle.
\end{equation}
We can see in Eq.\ (\ref{csfun}) that the analytic continuation $l \to \lambda$ does not pose any technical problem. 
The situation is the same with the Green's matrix. Both the matrix elements of $J$ and the $_{2}F_{1}$ are analytic in
terms of variables, so the analytic continuation amounts of straightforward substitutions $l \to \lambda$ and $Z \to Z'$
\cite{konya1999coulomb}.

\section{Extension to the Dirac equation}
\label{sec4}

The Dirac equation is a first order differential equation for the four-component wave function. 
Feynman and Gell-Mann ''squared'' it and obtained a second order differential equation for a
two-component wave function \cite{feynman1958theory}. 
If we assume that the potential term is associated only with the energy we have
\begin{equation} 
\left(\nabla^{2}+  \frac{1}{\hbar^{2}c^{2}} (E-V)^{2}  -  \frac{m^{2}c^{2}}{\hbar^{2}} \right) \psi + \frac{i}{\hbar c}   \grad V \cdot \vec{ \sigma} \psi = 0,
\label{eq:511101}
\end{equation}
where $\vec{\sigma}$ denotes the Pauli matrices.
If we separate off the rest energy, $E= E' + mc^{2}$, we get
\Beq
\left[ - \frac{\hbar^{2}}{2m} \nabla^{2}  + \tilde{V}     \right] \psi - 
 \frac{i \hbar c}{2mc^{2}}   \grad V \cdot \vec{ \sigma} \psi =  \varepsilon \psi.
\label{eq:511102}
\Eeq

Assume that the potential is spherical and it is a sum of a Coulomb plus short-range terms. 
Now the Hamiltonian forms a complete set of commuting observables with the total angular momentum operators 
$J^{2}$ and $J_{z}$. 
We follow the method of Ref.\ \cite{holstein2013topics}, with the difference that our formulae are valid even if the particle
is not charged.
The total wave function is a product radial and angular terms
\Beq
\psi = \frac{1}{r} \psi_{j}^{(\pm)}(r)  \Phi^{(\pm)}_{j,m} (\theta,\phi)~,
\Eeq
where
\Beq
\Phi^{(\pm)}_{j,m} (\theta,\phi) = \sum_{m_{l}, m_{s}} \langle l_{\pm} , 1/2 ; m_{l}, m_{s} | j, m \rangle Y_{l_{\pm}} (\theta,\phi)  \chi_{1/2, m_{s}}
\Eeq
are constructed by coupling orbital angular momentum $l_{\pm}$ to the spin such that $j = l_{+}+1/2 = l_{-}-1/2$. 
Then,
from Eq.\ (\ref{eq:511102}) we obtain the Hamiltonian
\Beq
\tilde{H} = - \frac{\hbar^{2}}{2m}  \pdv[2]{r} + \frac{\hbar^{2}}{2m} \frac{   l_{\pm}(l_{\pm}+1) - Z^{2} \alpha^{2} }{r^{2}}  
 - \frac{\hbar^{2}}{2m}    \frac{i  Z \alpha }{ r^{2}} ( \hat{\vec{r}}  \cdot \vec{ \sigma})   + \frac{Z'}{r}    +  w_{l}^{(s)}  
 + w'_{l} \left(  \hat{\vec{r}}  \cdot \vec{ \sigma}  \right) 
 \label{eq:511103}
\Eeq
where 
\Beq
w'_{l}(r)= -  \frac{i \hbar c}{2mc^{2}} \dv{v_{l}^{(s)} (r)}{r}.
\Eeq

We should recall that the parity operator ${\cal P}$, the mirroring of the coordinates, in polar coordinates, entails the transformation 
$\theta \to \pi -\theta$ and $\phi \to \phi+\pi$. 
The spherical harmonics transform as ${\cal P} \: Y_{lm} = (-)^{l} Y_{lm}$ and the electron has positive intrinsic parity. Consequently 
\begin{equation}
{\cal P} \:\Phi^{(\pm)}_{jm} = (-)^{l_{\pm}} \Phi^{(\pm)}_{jm},
\end{equation}
i.e.\   the states $\Phi^{(+)}_{jm}$ and $\Phi^{(-)}_{jm}$ have opposite parities.
We can also see that
\begin{equation}
\hat{\vec{r}} \cdot  \vec{ \sigma}     = \sigma_{x} \sin \theta \cos \phi +  \sigma_{y} \sin \theta \sin \phi + \sigma_{z} \cos \theta 
= \begin{pmatrix} \cos \theta & e^{-i\phi} \sin \theta \\   e^{i\phi} \sin \theta & -\cos \theta \end{pmatrix} 
\end{equation}
is an odd operator under parity, i.e.\ ${\cal P}   \hat{\vec{r}} \cdot  \vec{ \sigma}  = - \hat{\vec{r}} \cdot  \vec{ \sigma}$, 
and also $( \hat{\vec{r}} \cdot  \vec{ \sigma}    )^{2}=1$.  Additionally, we can easily verify by explicitly
calculating the commutator  that it commutes with ${\vec{J}}$ angular
momentum operator
\Beq
\comm{{\vec{J}}}{  \hat{\vec{r}} \cdot  \vec{ \sigma}    }= 0.
\Eeq
So,  $\hat{\vec{r}} \cdot \vec{ \sigma}$ acting 
 on $\ket{\Phi_{j,m}^{(\pm)}}$  does not change the eigenvalue $j$ .  
On the other hand,  $\hat{\vec{r}} \cdot \vec{ \sigma}$ is an odd operator
whose square is a unit operator. Its action on $\ket{\phi_{j,m}^{(\pm)}}$ should result in a state with 
opposite parity, i.e.\ $ \hat{\vec{r}} \cdot \vec{\sigma} $ should transform 
the states $\Phi^{(\pm)}_{j,m}$ into each other. So,  $(\hat{\vec{r}} \cdot \vec{ \sigma} )  \,\ket{ \Phi_{j,m}^{(\pm)}} = \ket{\Phi_{j,m}^{(\mp)}}$,
consequently $\mel{\Phi^{(\pm)}_{j,m} }{\hat{\vec{r}} \cdot \vec{ \sigma}  }{\Phi^{(\pm)}_{j,m} }=0$ and
$\mel{\Phi^{(\mp)}_{j,m} }{\hat{\vec{r}} \cdot \vec{ \sigma}    }{\Phi^{(\pm)}_{j,m} }=1$.

Eq.\ (\ref{eq:511103}) is a set of two-component coupled equations. The terms that are proportional to $\hbar^{2}/(2mr^{2})$ are given by
\begin{equation}
\begin{split}
 & \mel{ \Phi_{j,m}^{(\pm)}}{   l_{\pm}(l_{\pm}+1)   -Z^{2}\alpha^{2}-i Z \alpha \,{\vec \sigma}\cdot \hat{r} }{\Phi_{j,m}^{(\pm)}}  \\
  & =   
 \begin{pmatrix}  (j-1/2) (j+1/2) - Z^{2} \alpha^{2}  & -i Z \alpha \\ -i  Z \alpha & (j+1/2)(j+3/2) -Z^{2} \alpha^{2} \end{pmatrix}.
 \end{split}
\label{eql2z}
\end{equation}
By solving the matrix eigenvalue problem, we find the eigenstates $\ket{ \eta^{(\pm)} }$, which are  linear combinations of $\ket{ \Phi_{j,m}^{(\pm)} }$
\Beq
\mqty( \eta^{(+)} \\ \eta^{(-)}) = \frac{1}{2} \sqrt{\frac{j+1/2+s}{j+1/2}}  \mqty(  1 & i Z \alpha / (  j+1/2 + s  ) \\ 
  -i Z \alpha / (  j+1/2 + s) & 1 ) \mqty( \Phi^{(+)} \\ \Phi^{(-)}),
\Eeq
where 
\Beq
s = \sqrt{(j+1/2)^{2} -Z^{2}\alpha^{2}}.
\Eeq
We can equate the eigenvalues by $\lambda_{\pm}(\lambda_{\pm}+1)$ and find that 
\Beq
\lambda_{\pm}=s -1/2 \mp 1/2.
\Eeq
So,  
Eq.\ (\ref{eql2z}) in the $\ket{\eta^{(\pm)}}$ basis becomes diagonal
\begin{equation}
  \mel{ \chi_{j,m}^{(\pm)}}{   \hat{L}^{2}/\hbar^{2} -Z^{2}\alpha^{2}-i Z \alpha \,{\vec \sigma}\cdot \hat{r} }{\chi_{j,m}^{(\pm)}}  =   
 \begin{pmatrix}  \lambda_{+}(\lambda_{+}+1)     &  0  \\ 0 &  \lambda_{-}(\lambda_{-} +1) \end{pmatrix}.
\label{eql2z2}
\end{equation}
Consequently,  for Eq.\ (\ref{eq:511103}) we obtain
\Beq
  \left[ - \frac{\hbar^{2}}{2m}  \pdv[2]{r} +  \frac{\hbar^{2}}{2m}  \frac{ \lambda_{\pm}(\lambda_{\pm}+1 ) }{r^{2}}  
  + \frac{Z'}{r}    
 +   w_{l}^{(s)} + w'_{l}   \mqty( 0 &   1 \\ 1 & 0   )     \right]    \mqty(\psi_{j}^{(+)} \\ \psi_{j}^{(-)} )     
 =  \varepsilon  \mqty(\psi_{j}^{(+)} \\ \psi_{j}^{(-)} ).
\label{eq:511103b}
\Eeq
We can turn this differential equation into a Lippmann-Schwinger form
\Beq
 \mqty(\psi_{j}^{(+)} \\ \psi_{j}^{(-)} ) = 
 \mqty( g^{C}_{\lambda_{+}} (Z',\varepsilon) &  0 \\ 0 & g^{C}_{\lambda_{-}} (Z',\varepsilon) )   
 \mqty(  \underline{\tilde{w}}_{\lambda_{+}, \lambda_{+}} &   \underline{\tilde{w}}_{\lambda_{+}, \lambda_{-}} \\
 \underline{\tilde{w}}_{\lambda_{-}, \lambda_{+}} &   \underline{\tilde{w}}_{\lambda_{-}, \lambda_{-}}  )
 \mqty(\psi_{j}^{(+)} \\ \psi_{j}^{(-)} ),
\Eeq
where $\underline{\tilde{w}}_{\lambda_{\pm},\lambda_{\pm}} = 
\langle{n \lambda_{\pm}} | w_{l}^{(s)} | n' \lambda_{\pm} \rangle$ and
$\underline{\tilde{w}}_{\lambda_{\pm},\lambda_{\mp}} = 
\langle{n \lambda_{\pm}} | w'_{l} | n' \lambda_{\mp} \rangle$. 
Then, the energy $\varepsilon$  can be determined by the zeros of the determinant
\Beq
\mqty| (g^{C}_{\lambda_{+}} (Z',\varepsilon))^{-1}  -  \underline{\tilde{w}}_{\lambda_{+}, \lambda_{+}}   
 &  - \underline{\tilde{w}}_{\lambda_{+}, \lambda_{-}}   \\   -\underline{\tilde{w}}_{\lambda_{-}, \lambda_{+}}  
 & (g^{C}_{\lambda_{-}} (Z',\varepsilon) )^{-1}    -  \underline{\tilde{w}}_{\lambda_{-}, \lambda_{-}}   | =0.
\Eeq
The corresponding matrix elements can be calculated the same way as before, by performing an analytic continuation
in the non-relativistic formulae $l \to \lambda_{\pm}$ and $Z\to Z'$.

\section{Numerical Illustrations}
\label{sec5}

As numerical illustrations we consider the model with $m=1$, $\hbar=1$, $e^{2} =1$ and $\alpha= e^{2}/\hbar c  = 1/137.03604$.
We take $Z=50$ and 
\Beq
v_{l}^{(s)}(r) = -240 \exp(-r)/r + 320 \exp(-4r)/r. 
\Eeq
The Schr\"odinger, the Klein-Gordon and Dirac bound and resonant state results for $l=0,1,2$ are given in Table \ref{table1}.
The complex energies are given by $E=E_{r}-i \Gamma/2$, 
where $E_{r}$ is the resonance energy and $\Gamma/2$ is the lifetime of the
resonant state. 

\begin{table}
\centering
\caption{Non-relativistic energies.  }
\label{table1}

\begin{tabular}{| l  | l | l  |   }
\hline
 \multicolumn{3}{|c|}{Non-relativistic energies} \\
\hline
 l=0 &  l=1  &  l=2   \\
\hline
 -92.264199  &-86.36494   &    -75.76312	   	  \\
 -54.224609    &-49.69048  &     -41.60855 	    \\
 -26.210528   &-22.84595 &   -16.91107  	    \\
 -6.5302229    &-4.175213  &     -0.091564	    \\
 6.139886      &   7.600636     &   10.019283  	    \\
   \,\,\,\,\,-0.00000002 $\textbf{\textit{i}}$    &     \,\,\,\,\, -0.0000000003 $\textbf{\textit{i}}$ &   \,\,\,\,\,-0.000017 $\textbf{\textit{i}}$	    \\

  \hline
   \multicolumn{3}{|c|}{Klein-Gordon energies} \\
\hline
-92.27553   &  -86.37913  &   -75.78641	  \\
-54.25825   &  -49.72854 &	-41.65660   \\
-26.26132   &   -22.89969 &	-16.97094   \\
-6.583528  &   -4.229067  &	  -0.146437 \\
  6.098560    &     7.561164  	&   10.019284\\
   \,\,\,\,\,  -0.000000003 $\textbf{\textit{i}}$  &    \,\,\,\,\,  -0.0000000002 $\textbf{\textit{i}}$	& \,\,\,\,\,  -0.000014 $\textbf{\textit{i}}$ \\

\hline
\multicolumn{3}{|c|}{Dirac energies} \\
\hline
    j=1/2 &  j=3/2  &  j=5/2   \\
\hline
 -91.73292 &  -86.20067 & -75.66290 \\
 -86.80452 & -75.92108 & -62.28595 \\
 -54.03927 & -49.62861 & -41.59556 \\
 -49.89311 & -41.72801 & -31.45510 \\
 -26.13929 & -22.84399 &  -16.94087 \\
   -22.98923& -17.01021 & -9.587586  \\
 -6.503713 & -4.195174 & -0.130783 \\
 -4.285842&  -0.169263 & 4.786557 \\
6.1480973 &   7.580558   & \,\,\,\,\, -0.0000000000006 $\textbf{\textit{i}}$	   \\
   \,\,\,\,\,  -0.00000008 $\textbf{\textit{i}}$	&   \,\,\,\,\,  -0.0000000004 $\textbf{\textit{i}}$	 & 9.991357  \\
  & 9.972041   &  \,\,\,\,\,-0.000015 $\textbf{\textit{i}}$	   \\
 \,\,\,\,\,   &\,\,\,\,\,-0.000014 $\textbf{\textit{i}}$	  &   \\
   \hline
\end{tabular} 

\end{table}

\section{Summary and Conclusions}
\label{sec6}

In this work we have extended a quantum mechanical approximation method that has been rather successful in 
non-relativistic calculations to calculate bound and resonant states of the relativistic Klein-Gordon and Dirac 
equations. We brought the relativistic equations in a form similar to the non-relativistic Schr\"odinger equation.
We accomplish this by redefining the angular momentum $l \to \lambda$, the charge $Z \to Z'$, 
the energy $E \to \varepsilon$, the short-range potential $v^{(s)}  \to w^{(s)}$ and the Green's operator
$g_{l}^{C}(Z,E) \to g_{l}^{C}(Z',\varepsilon)$. 
This way all the advantages of the method have been retained and transferred to study relativistic problems.


\bibliographystyle{spphys}       
\bibliography{fv00}

\begin{thebibliography}{10}
\providecommand{\url}[1]{{#1}}
\providecommand{\urlprefix}{URL }
\expandafter\ifx\csname urlstyle\endcsname\relax
  \providecommand{\doi}[1]{DOI \discretionary{}{}{}#1}\else
  \providecommand{\doi}{DOI \discretionary{}{}{}\begingroup
  \urlstyle{rm}\Url}\fi

\bibitem{papp1987bound}
Z.~Papp, Journal of Physics A: Mathematical and General \textbf{20}(1), 153
  (1987)

\bibitem{papp1988potential}
Z.~Papp, Physical Review C \textbf{38}(5), 2457 (1988)

\bibitem{PhysRevC.61.034302}
B.~K{\'o}nya, G.~L{\'e}vai, Z.~Papp, Phys. Rev. C \textbf{61}, 034302 (2000)

\bibitem{PhysRevC.54.50}
Z.~Papp, W.~Plessas, Phys. Rev. C \textbf{54}(1), 50 (1996)

\bibitem{PhysRevC.55.1080}
Z.~Papp, Phys. Rev. C \textbf{55}, 1080 (1997)

\bibitem{PhysRevA.63.062721}
Z.~Papp, C.Y. Hu, Z.T. Hlousek, B.~K\'onya, S.~Yakovlev, Phys. Rev. A
  \textbf{63}(6), 062721 (2001)

\bibitem{brown2013approximations}
N.C. Brown, S.E. Grefe, Z.~Papp, Physical Review C \textbf{88}(4), 047001
  (2013)

\bibitem{Konya:1997JMP}
B.~K{\'o}nya, G.~L{\'e}vai, Z.~Papp, J.Math.Phys. \textbf{38}, 4832 (1997)

\bibitem{PRADemir2006}
F.~Demir, Z.T. Hlousek, Z.~Papp, Phys. Rev. A \textbf{74}, 014701 (2006)

\bibitem{baker1975essentials}
G.~Baker~Jr, \emph{Essentials of Pade approximants} (Academic Press, New York,
  1975)

\bibitem{PhysRevA.46.4437}
Z.~Papp, Phys. Rev. A \textbf{46}, 4437 (1992)

\bibitem{konya1999coulomb}
B.~K{\'o}nya, Z.~Papp, Journal of Mathematical Physics \textbf{40}(5), 2307
  (1999)

\bibitem{feynman1958theory}
R.P. Feynman, M.~Gell-Mann, Physical Review \textbf{109}(1), 193 (1958)

\bibitem{holstein2013topics}
B.R. Holstein, \emph{Topics in advanced quantum mechanics} (Courier
  Corporation, 2013)




\end{thebibliography}

%
%

\end{document}